\documentclass[12pt]{article}

\usepackage{graphicx}
\begin{document}
\begin{center}
{\bf On Generalized Logarithmic Electrodynamics}\\
\vspace{5mm}
 S. I. Kruglov
\footnote{E-mail: serguei.krouglov@utoronto.ca}
 \\

\vspace{5mm}
\textit{Department of Chemical and Physical Sciences, University of Toronto,\\
3359 Mississauga Road North, Mississauga, Ontario, Canada L5L 1C6}
\end{center}

\begin{abstract}

The generalized logarithmic electrodynamics with two parameters $\beta$ and $\gamma$ is considered. The indexes of refraction of light in the external magnetic field are calculated. In the case $\beta=\gamma$ we come to results obtained by P. Gaete and J. Helay\"{e}l-Neto \cite{Gaete} (Eur.Phys.J. C\textbf{74}, 2816 (2014)). The bound on the values of $\beta$, $\gamma$ was obtained from the Bir\'{e}fringence Magn\'{e}tique du Vide (BMV) experiment. The symmetrical Belinfante energy-momentum tensor and dilatation current are found.

\end{abstract}

\section{Introduction}

Nonlinear classical electrodynamics in vacuum is of interest because of the one-loop quantum corrections in QED which give non-linear terms \cite{Heisenberg}, \cite{Schwinger}. In addition, to solve the problem of singularity of a point-like charge giving an infinite electromagnetic energy, Born and Infeld (BI) \cite{Born}, \cite{Infeld} introduced a new parameter with the dimension of length. BI non-linear electrodynamics results in a finite electromagnetic energy of point-like particles.
Other examples of non-linear electrodynamics were introduced in \cite{Kruglov}, \cite{Kruglov3}, \cite{Gaete}, \cite{Hendi} and \cite{Kruglov4}. In the vacuum, in the presence of strong external magnetic field, non-linear effects can be observed in experiments. Thus, PVLAS \cite{PVLAS} and BMV \cite{Rizzo} experiments can give the bounds on dimensional parameters introduced in non-linear electrodynamics.
In this letter we calculate the values of indexes of refraction of light in the external magnetic field in generalized logarithmic electrodynamics and estimate the bound on the values of the parameters $\beta$, $\gamma$ from the BMV experiment.

The Heaviside-Lorentz system with $\hbar =c=\varepsilon_0=\mu_0=1$ and Euclidian metric are explored.

\section{The model of generalized logarithmic electrodynamics}

Let us consider the Lagrangian density of non-linear generalized logarithmic electrodynamics
\begin{equation}
{\cal L} = -\beta^2\ln\left(1+\frac{{\cal F}}{\beta^2}- \frac{{\cal G}^2}{2\beta^2\gamma^2}\right),
 \label{1}
\end{equation}
where $\beta$, $\gamma$ are dimensional parameters and ${\cal F}=(1/4)F_{\mu\nu}^2=(\textbf{B}^2-\textbf{E}^2)/2$, ${\cal G}=(1/4)F_{\mu\nu}\tilde{F}_{\mu\nu}=\textbf{E}\cdot \textbf{B}$ are the Lorentz-invariants. The field strength is $F_{\mu\nu}=\partial_\mu A_\nu-\partial_\nu A_\mu$, and $\tilde{F}_{\mu\nu}=(1/2)\varepsilon_{\mu\nu\alpha\beta}F_{\alpha\beta}$ is a dual tensor ($\varepsilon_{1234}=-i$).
At $\gamma=\beta$ we arrive at logarithmic electrodynamics considered in \cite{Gaete}.

We obtain from Eq. (1), and the Euler-Lagrange equations, the equations of motion
\begin{equation}
\partial_\mu\left[\frac{1}{\Lambda}\left(F_{\mu\nu}-\frac{{\cal G}\tilde{F}_{\mu\nu}}{\gamma^2}\right) \right]=0,
\label{2}
\end{equation}
where
\begin{equation}
\Lambda=1+\frac{{\cal F}}{\beta^2}- \frac{{\cal G}^2}{2\beta^2\gamma^2}.
\label{3}
\end{equation}
From Eq. (1) and the expression for the electric displacement field
$\textbf{D}=\partial{\cal L}/\partial \textbf{E}$ ($E_j=iF_{j4}$), one finds
\begin{equation}
\textbf{D}=\frac{1}{\Lambda}\left[ \textbf{E}+\frac{(\textbf{E}\cdot\textbf{B})\textbf{B}}{\gamma^2}\right].
\label{4}
\end{equation}
Defining the tensor of the electric permittivity $\varepsilon_{ij}$ by the relation $D_i=\varepsilon_{ij}E_j$, we obtain
\begin{equation}
\varepsilon_{ij}=\frac{1}{\Lambda}\left[\delta_{ij}+\frac{1}{\gamma^2} B_iB_j\right].
\label{5}
\end{equation}
Using the definition of the magnetic field $\textbf{H}=-\partial{\cal L}/\partial \textbf{B}$ ($B_j=(1/2)\varepsilon_{jik}F_{ik}$) and Eq. (1), one finds
\begin{equation}
\textbf{H}= \frac{1}{\Lambda}\left[\textbf{B}-\frac{(\textbf{E}\cdot\textbf{B})\textbf{E}}{\gamma^2}\right].
\label{6}
\end{equation}
Introducing the magnetic induction field $B_i=\mu_{ij}H_j$, the inverse magnetic permeability tensor $(\mu^{-1})_{ij}$ is given by
\begin{equation}
(\mu^{-1})_{ij}=\frac{1}{\Lambda}\left[\delta_{ij}-\frac{1}{\gamma^2} E_iE_j\right].
\label{7}
\end{equation}
From the field equations (2) and the Bianchi identity $\partial_\mu \widetilde{F}_{\mu\nu}=0$, one can write Maxwell's equations as follows:
\[
\nabla\cdot \textbf{D}= 0,~~~~ \frac{\partial\textbf{D}}{\partial
t}-\nabla\times\textbf{H}=0.
\]
\begin{equation}
\nabla\cdot \textbf{B}= 0,~~~~ \frac{\partial\textbf{B}}{\partial
t}+\nabla\times\textbf{E}=0,
\label{8}
\end{equation}
where the electric permittivity $\varepsilon_{ij}$ and magnetic permeability $\mu_{ij}$ depend on the fields $\textbf{E}$ and $\textbf{B}$ and are given by Eqs. (5),(7).

\section{Vacuum birefringence}

Let us consider the presence of the external constant and uniform magnetic induction field $\textbf{B}_0=(B_0,0,0)$ and the plane electromagnetic wave, $(\textbf{e}, \textbf{b}$),
\begin{equation}
\textbf{e}=\textbf{e}_0\exp\left[-i\left(\omega t-kz\right)\right],~~~\textbf{b}=\textbf{b}_0\exp\left[-i\left(\omega t-kz\right)\right].
\label{9}
\end{equation}
which propagates in the $z$-direction.
As a result, the electromagnetic fields become $\textbf{E}=\textbf{e}$, $\textbf{B}=\textbf{b}+\textbf{B}_0$. We investigate the case when amplitudes of electromagnetic wave $e_0, b_0$ are smaller comparing with the strong magnetic induction field, $e_0,b_0\ll B_0$. In this approximation, up to ${\cal O}(e_0^2)$, ${\cal O}(b_0^2)$, the Lagrangian density (1) is
\begin{equation}
{\cal L} \approx -\beta^2\ln\left[1+\frac{(\textbf{B}_0+\textbf{b})^2-\textbf{e}^2}{2\beta^2}- \frac{(\textbf{e}\cdot\textbf{B}_0)^2}{2\beta^2\gamma^2}\right].
 \label{10}
\end{equation}
Defining the fields \cite{Dittrich} $d_i=\partial{\cal L}/\partial e_i$, $h_i=-\partial{\cal L}/\partial b_i$ and
linearizing these equations with respect to the wave fields $\textbf{e}$ and $\textbf{b}$, we obtain the electric permittivity and magnetic permeability tensors
\begin{equation}
\varepsilon_{ij}=\frac{1}{\lambda}\left(\delta_{ij}+\frac{1}{\gamma^2} B_{0i}B_{0j}\right),~
(\mu^{-1})_{ij}=\frac{1}{\lambda}\left(\delta_{ij}-\frac{B_{0i}B_{0j}}{\lambda\beta^2}\right),~
\lambda=1+\frac{\textbf{B}_0^2}{2\beta^2}.
\label{11}
\end{equation}
The electric permittivity and magnetic permeability tensors are diagonal with the elements
\begin{equation}
\varepsilon_{11}=\frac{1}{\lambda}\left(1+\frac{B_0^2}{\gamma^2}\right),~~\varepsilon_{22}=\varepsilon_{33}
=\frac{1}{\lambda},~~\mu_{11}=\frac{\lambda}{\left(1-\frac{B_0^2}{\lambda\beta^2}\right)},~~\mu_{22}=\mu_{22}=\lambda.
\label{12}
\end{equation}
If the polarization is parallel to external magnetic field, $\textbf{e}=e_0(1,0,0)$, and one finds from Maxwell's equations that $\mu_{22}\varepsilon_{11}\omega^2=k^2$. As a result, the index of refraction is given by
\begin{equation}
n_\|=\sqrt{\mu_{22}\varepsilon_{11}}=\sqrt{1+\frac{B_0^2}{\gamma^2}}.
\label{13}
\end{equation}
In the case when the polarization of the electromagnetic wave is perpendicular to external induction magnetic field, $\textbf{e}=e_0(0,1,0)$, and $\mu_{11}\varepsilon_{22}\omega^2=k^2$. The index of refraction is
\begin{equation}
n_\perp=\sqrt{\mu_{11}\varepsilon_{22}}=\sqrt{\frac{1+B_0^2/(2\beta^2)}{1-B_0^2/(2\beta^2)}}.
\label{14}
\end{equation}
At $\beta=\gamma$ we arrive at the expressions for $n_\|$, $n_\perp$ obtained in \cite{Gaete}.
The phase velocity depends on the polarization of the electromagnetic wave.
With the help of the approximation $ B_0^2/\beta^2\ll 1$, $ B_0^2/\gamma^2\ll 1$ we find from Eqs. (13),(14)
\[
n_\|\approx 1+\frac{B_0^2}{2\gamma^2},~~~n_\perp \approx 1+\frac{B_0^2}{2\beta^2}.
\]
At the case $\beta=\gamma$, one has $n_\|=n_\perp$ and the effect of birefringence disappears.
But in QED with one loop corrections, we have the relation $n_\|>n_\perp$ \cite{Dittrich}. Therefore in the generalized logarithmic electrodynamics with two parameters which we introduced the phenomenon of birefringence takes place if $\gamma<\beta$  ($n_\|>n_\perp$).

In the presence of a transverse external magnetic field the phenomenon of birefringence is named the Cotton-Mouton (CM) effect \cite{Battesti}. The difference in indexes of refraction is defined by the relation
\begin{equation}
\triangle n_{CM}=n_\|-n_\perp=k_{CM}B_0^2.
\label{15}
\end{equation}
From Eqs. (13),(14) at $ B_0^2/\beta^2\ll 1$, $ B_0^2/\gamma^2\ll 1$ one finds
\begin{equation}
\triangle n_{CM}\approx \frac{B_0^2}{2}\left(\frac{1}{\gamma^2}-\frac{1}{\beta^2}\right),
\label{16}
\end{equation}
and CM coefficient becomes
\begin{equation}
k_{CM}\approx \frac{1}{2}\left(\frac{1}{\gamma^2}-\frac{1}{\beta^2}\right).
\label{17}
\end{equation}
The vacuum magnetic linear birefringence by the BMV experiment for a maximum field of $B_0=6.5$ T
gives the value \cite{Rizzo}
\begin{equation}
k_{CM}=(5.1\pm 6.2)\times 10^{-21} \mbox {T}^{-2}.
\label{18}
\end{equation}
From Eqs. (17),(18), we obtain the bound on the parameters $\beta$, $\gamma$:
\begin{equation}
\frac{1}{\gamma^2}-\frac{1}{\beta^2} \approx 10^{-20} \mbox {T}.
\label{19}
\end{equation}
We note that the value obtained from QED, using one loop approximation, is \cite{Rizzo} $k_{CM}\approx 4.0\times 10^{-24} \mbox {T}^{-2}$ which is much less than the experimental value (18).

\section{The energy-momentum tensor and dilatation current}

From Eq. (1) using the method of \cite{Coleman}, we obtain the symmetrical
Belinfante tensor
\begin{equation}
T_{\mu\nu}^{B}=-\frac{1}{\Lambda}F_{\nu\alpha}\left(F_{\mu\alpha}-\frac{{\cal G}\tilde{F}_{\mu\alpha}}{\gamma^2}\right)-\delta_{\mu\nu}{\cal L},
\label{20}
\end{equation}
where $\Lambda$ is given by Eq. (3). The energy density found from Eq. (20) is \footnote{There is a typo in \cite{Gaete} (Eq. (26))}
\begin{equation}
T_{44}^{B}=\frac{1}{\Lambda}\left(\textbf{E}^2-\frac{{\cal G}^2}{\gamma^2}\right)+\beta^2\ln\Lambda.
\label{21}
\end{equation}
The trace of the energy-momentum tensor (20) is as follows:
\begin{equation}
T_{\mu\mu}^{B}=-\frac{4}{\Lambda}\left({\cal F}-\frac{{\cal G}^2}{\gamma^2}\right)+4\beta^2\ln\Lambda,
\label{22}
\end{equation}
and the trace of the energy-momentum tensor is not zero contrarily to classical electrodynamics. According to \cite{Coleman}, we obtain the dilatation current (the field-virial $V_\mu$ is zero)
\begin{equation}
D_{\mu}^{B}=x_\alpha T_{\mu\alpha}^{B},
\label{23}
\end{equation}
and the divergence of dilatation current is
\begin{equation}
\partial_\mu D_{\mu}^{B}=T_{\mu\mu}^B.
\label{24}
\end{equation}
As a result, the scale (dilatation) symmetry is broken because the dimensional parameters $\beta$, $\gamma$ were introduced. In BI electrodynamics the dilatation symmetry is also broken \cite{Kruglov5} in opposite to the linear Maxwell electrodynamics.

\section{Conclusion}

We have considered the model of generalized logarithmic electrodynamics with two parameters $\beta$ and $\gamma$. At
$\beta=\gamma$ we arrive at logarithmic electrodynamics considered in \cite{Gaete}. We show that at $\beta=\gamma$ in the approximation $ B_0^2/\beta^2\ll 1$, $ B_0^2/\gamma^2\ll 1$ the phenomenon of birefringence is absent. Classical electrodynamics with QED corrections gives the effect of birefringence and, therefore, the case $\beta\neq\gamma$ is important.
From the BMV experiment we have obtained the bound on the parameters $\beta$ and $\gamma$. Another bound on the parameter $\beta$ from the point of view that the electron mass has pure electromagnetic nature was proposed in \cite{Gitman}. The scale symmetry is broken and dilatation current was found in the model under consideration.

\vspace{5mm}
\textbf{Acknowledgments}
\vspace{5mm}

I wish to thank P. Gaete and J. Helay\"{e}l-Neto for communications that allowed me to make corrections in the first version of this letter.

\end{document}